\documentclass[lettersize,journal]{IEEEtran}
\usepackage{amsmath,amsfonts}
\usepackage{booktabs}
\usepackage{algorithm}
\usepackage{algpseudocode}  % 或 algorithmicx
\usepackage{array}
\usepackage{xcolor}
\usepackage{soul}
\soulregister\cite7
\sethlcolor{yellow}
\usepackage{textcomp}
\usepackage{multirow}
\usepackage{stfloats}
\usepackage{url}
\usepackage{verbatim}
\usepackage{graphicx}
\usepackage{balance}
\usepackage{stfloats}
\usepackage{cite}
\usepackage{graphicx}
\usepackage{amsmath}
\usepackage{amssymb}
\usepackage{bm} % for bold math symbols if needed
\usepackage{pifont}
\usepackage{adjustbox}
\usepackage{enumitem}
\begin{document}

\title{Learned Score-Based Turbo Message Passing for Compressive Image Recovery}

\author{Peihang Li, Hao Jiang,~\IEEEmembership{Graduate Student Member,~IEEE,} and Xiaojun Yuan,~\IEEEmembership{Fellow,~IEEE}
        % <-this % stops a space
\thanks{Peihang Li, Hao Jiang and Xiaojun Yuan are with the National Key Laboratory of Wireless Communications, University of Electronic Science and Technology of China, Chengdu 611731, China (e-mail: 202621220232@std.uestc.edu.cn;jh@std.uestc.edu.cn; xjyuan@uestc.edu.cn).}% <-this % stops a space
}

\markboth{}%
{Shell \MakeLowercase{\textit{et al.}}: A Sample Article Using IEEEtran.cls for IEEE Journals}


\maketitle
\newcommand{\pri}{\mathrm{pri}}
\newcommand{\post}{{post}}
\newcommand{\ext}{{ext}}
% 重定义 Require 和 Ensure
\renewcommand{\algorithmicrequire}{\textbf{Input:}}
\renewcommand{\algorithmicensure}{\textbf{Output:}}

\begin{abstract}
Score-based generative models have been incorporated into message passing algorithms for compressive image recovery, providing a prior for the denoising module in message passing. Score-based generative models assume the input signal to be a clean signal perturbed by independent and identically distributed (i.i.d.) Gaussian noise. Yet, this assumption often fails during message passing, leading to slow convergence and limited reconstruction quality. In this letter, we propose the learned score-based turbo message passing (LSTMP) algorithm to overcome this limitation. By employing a deep unfolding framework, LSTMP unrolls the iterative process into a neural network, enabling the denoiser to adapt to complicated noise distributions inherent in the iterative updates. Furthermore, by treating the denoiser as an iterative solver, we incorporate the likelihood information of compressed measurements into the neural network of the denoiser, resembling the message passing between linear and nonlinear modules in STMP. Experiments on the FFHQ dataset show that LSTMP achieves convergence in no more than three iterations while delivering high-fidelity reconstructions, significantly outperforming the existing baselines.
% By integrating the score-based generative model, message passing algorithms have been applied in compressive image recovery. The score-based network is applied in the denoising module of such algorithms to provide effective prior knowledge. However, score-based generative models assume the input signal to be a signal perturbed by i.i.d. Gaussian noise, which significantly limits the convergence speed and reconstruction quality. Recently, deep unfolding networks have been widely adopted in compressed sensing due to their suitability for iterative algorithms. In this paper, we introduce the learned score-based turbo message passing (LSTMP). LSTMP applies a deep unfolding framework to the score-based message passing algorithm, unrolling the iterative process into a neural network. It enables the network to effectively handle a broader range of signals, rather than being limited to those perturbed by i.i.d. Gaussian noise. To further accelerate convergence, we incorporate a residual denoising network and the turbo-guided linear correction (TGLC) module into the framework. Experiments on the FFHQ dataset show that LSTMP achieves convergence in approximately 3 iterations while delivering high-fidelity reconstructions.
\end{abstract}


\section{Introduction}
\IEEEPARstart{C}{ompressive} image recovery\cite{compressive_image_recovery} is the process of reconstructing an image from highly undersampled measurements and has been widely applied in numerous imaging fields, including magnetic resonance imaging (MRI)\cite{MRI}, computed tomography (CT)\cite{CT} and single-pixel camera (SPI)\cite{SPI}. In these fields, hardware constraints often restrict the number of measurements to be significantly smaller than the signal dimension, which makes it necessary to provide effective prior information for the reconstruction process.

Message passing algorithms are regarded as effective approaches for solving such compressed sensing problems due to their computational efficiency and theoretical guarantees. Prominent examples include approximate message passing (AMP)\cite{amp}, turbo message passing (TMP)\cite{turbo-cs}, denoising-based turbo compressed sensing (D-Turbo-CS)\cite{XJAZ03B41BBF891512CD912CDF1605452A7F} and denoising-based AMP (D-AMP). Traditionally, image priors in these algorithms are derived from either hand-crafted rules or off-the-shelf denoisers. However, such priors are often insufficient to describe the complex statistical structures of natural images, limiting reconstruction performance. 
To better capture complex image priors, methods like LAMP\cite{2017AMP} and LTMP\cite{2019Learned} employ deep unfolding\cite{deep_unfold1, deep_unfold2, lista} to transform traditional iterative algorithms into trainable deep networks. Nevertheless, their ability to exploit complex image priors remains limited by the design of their denoisers.
To address this limitation, the score-based generative models\cite{2020Score} have been incorporated into message passing frameworks. As a pioneering example, the score-based turbo message passing (STMP) \cite{2025Score} algorithm iterates between a linear module (that handles the likelihood information provided by the measurements) and a non-linear denoiser (that exploits the image prior via a pre-trained generative model), yielding comparable image reconstruction performance but with much lower computational complexity, as compared to its diffusion-based counterparts\cite{dmps, ddrm, diffpir, diffusion3}.
% The score-based generative model is employed to estimate the corresponding score function. It is then used in the denoising module of the message passing algorithm to provide effective prior knowledge for image recovery. 
Nevertheless, in STMP, the input to the denoiser is assumed to be an additive white Gaussian noise (AWGN) observation of the ground-truth. This idealized assumption does not hold in many practical scenarios, which leads to limited reconstruction performance and high computational complexity.

In this letter, we propose a learning-based algorithm named learned score-based turbo message passing (LSTMP) for compressive image recovery. First, following the idea of deep unfolding, LSTMP unrolls the iterative inference procedure of STMP into a trainable neural network. A key advantage of this design is that the denoiser can adapt to complicated error distributions that naturally arise during the message passing iterations, rather than requiring the input to be modelled as an AWGN observation of the ground-truth. Second, we further treat the neural network used in the denoiser as an iterative solver for image reconstruction, referred to as \textit{refolding}. 
% Specifically, the TGLC module is a lightweight linear block that incorporates physical measurement information into the denoising process. The embedding of TGLC in the denoiser enables the denoising network to better exploit the likelihood constraint, thereby improving its denoising capability. 
By integrating lightweight linear modules for absorbing likelihood information, the refolding operation (i.e., the denoiser) resembles the iterative process between the linear and non-linear modules of STMP. 
Notably, the linear modules introduce only 0.127M additional parameters in each layer.
Extensive experiments on the FFHQ dataset demonstrate that LSTMP, by incorporating the ideas of unfolding and refolding, achieves rapid convergence while maintaining high reconstruction fidelity, showing significant advantages over STMP and the diffusion-based algorithms\cite{dmps, ddrm, diffpir} in terms of performance-complexity tradeoff.
% \begin{enumerate}[label=(\arabic*)]
%     \item We propose LSTMP, a turbo message passing algorithm that integrates deep unfolding framework and score-based priors for compressive image recovery.
%     \item We develop TGLC modules for the denoising network. The TGLC module guides the update of the denoising network by introducing the likelihood constraint, aligning internal denoising process of the network with the turbo message passing procedure.
%     \item Extensive experiments on the FFHQ dataset demonstrate that LSTMP achieves rapid convergence while maintaining high reconstruction fidelity, showing significant advantages over STMP and many diffusion-based algorithms. It is worth noting that LSTMP requires only about 3 iterations to converge.
    
% \end{enumerate}

% In this letter, we propose a learned score-based turbo message passing (LSTMP) algorithm which integrates the STMP algorithm with the deep unfolding framework. Unlike standard STMP, LSTMP fine-tunes the network in the denoiser, enabling it to adapt to complex noise distributions rather than relying on idealized AWGN assumptions. To accelerate convergence, we further incorporate a residual denoising network and turbo-guided linear correction (TGLC) modules within this framework. Extensive experiments on the FFHQ dataset demonstrate that LSTMP achieves rapid convergence while maintaining high reconstruction fidelity, showing significant advantages over STMP and many diffusion-based algorithms.

\section{Preliminaries}
\label{sec:section2}
\subsection{System Model}
Consider the following linear inverse system
\begin{equation}
    \bm{y}=\bm{Ax}+\bm{n},
    \label{inverse problem}
\end{equation}
where $\bm{x} \in \mathbb{R}^N$ is an image signal, $\bm{A}\in \mathbb{R}^{M \times N}$ is a known measurement matrix, $\bm{y} \in \mathbb{R}^M$ denotes the observed image signal, and $\bm{n} \sim \mathcal{N}(\bm{0}, \delta^2 \bm{I})$ refers to AWGN. Due to physical or hardware limitations, the measurement matrix $\bm{A}$ in \eqref{inverse problem} typically satisfies the condition $M<N$ and is partial orthogonal (i.e., $\bm A \bm A^\top=\bm I$). We aim to recover the original image signal $\bm{x}$ from the observed signal $\bm{y}$.

\subsection{STMP Framework}

STMP\cite{2025Score} provides a new method to recover the true signal $\bm x$ from limited observations $\bm y$ in \eqref{inverse problem}. As shown in Algorithm \ref{alg:stmp}, STMP consists of two modules: the linear minimum mean-squared error (LMMSE) estimator and the score-based minimum mean-squared error (MMSE) denoiser. Leveraging the partial orthogonality of $\bm A$ allows for an efficient update of the posterior statistics. Given $\bm x_A^{pri}$ and $v_A^{pri}$ as the prior information, the posterior mean $\bm x_A^{post}$ is obtained via LMMSE estimation\cite{signal_process}, which is given by
\begin{equation}
    \bm{x}_{A}^\post = \bm{x}_{A}^\pri + \frac{v_{A}^\pri}{v_{A}^\pri + \delta^2} \bm{A}^\top (\bm{y} - \bm{A}\bm{x}_{A}^\pri).
    \label{lmmse}
\end{equation}
The posterior variance $v_A^{post}$ is updated in Line 4. In Lines 5 and 6, the posterior information is converted into extrinsic messages\cite{extrinsic_message} $\bm{x}_A^{ext}$ and ${v}_A^{ext}$ by excluding the contribution of the input message, and the latter then serve as the input of the score-based MMSE denoiser.

%In each iteration, LMMSE estimator uses the observed data $\bm{y}$ and prior information from the score-based MMSE denoiser to compute the posterior mean $\bm{x}_A^{post}$ and posterior variance $v_A^{post}$. They are then converted into extrinsic messages\cite{extrinsic_message} $\bm{x}_A^{ext}$ and ${v}_A^{ext}$ by excluding the contribution of the input message. The extrinsic messages serves as the input to the score-based MMSE denoiser.

In the score-based MMSE denoiser, STMP models the prior information $\bm{x}_B^{pri}$ as an AWGN observation of the ground-truth signal $\bm{x}$: 
\begin{equation}
    \bm{x}_B^{pri}=\bm{x}+\sigma \bm{n} \quad \bm{n} \sim \mathcal{N}\left(0,  I\right),
    \label{xb model}
\end{equation} 
where $\sigma$ is the prior standard deviation of the MMSE denoiser, i.e., $\sigma=\sqrt{v_B^{pri}}$. As such, in Line 9, the posterior mean $\bm{x}_{B}^\post$ is given by Tweedie's formula\cite{2011Tweedie}, 
% \begin{equation}
%     \bm{x}_{B}^\post = \bm{x}_{B}^\pri + v_{B}^\pri \nabla_{{\bm{x}}_B^{pri}} \log p({{\bm{x}}_B^{pri}}).
% \end{equation}
% \begin{equation}
%     v_{B}^\post = v_{B}^\pri + {(v_{B}^\pri)^2} \nabla^2_{{\bm{x}}_B^{pri}} \log p({{\bm{x}}_B^{pri}}).
% \end{equation}
where the score model $\bm{s_{\theta}}(\cdot)$ parameterized with $\bm \theta$ is learned to match the score function $\nabla_{{\bm{x}}_B^{pri}} \log p({{\bm{x}}_B^{pri}})$. The posterior variance $v_B^{post}$ is given by Line 10. 
% This design enables STMP to exploit effective, data-driven image priors.
The posterior mean and variance are then converted to extrinsic messages and fed back into the LMMSE estimator as prior information from Lines 11 to Line 13. This summarizes the full iterative process of STMP.

\begin{algorithm}[t]
\caption{STMP Algorithm}
\label{alg:stmp}
\begin{algorithmic}[1]
% \setstretch{0.8} 
\Require $\bm{A}$, $\bm{y}$, $\delta^2$, $\bm{x}_\mathrm{A}^{pri}$, $v_\mathrm{A}^{pri}$
\Ensure $\bm{x}_\mathrm{B}^\post$
\Repeat
    \State \text{\% LMMSE estimator}
    \vspace{-2pt}
    % \State $\bm{x}_{A}^\post = \bm{x}_{A}^\pri + \frac{v_{A}^\pri}{v_{A}^\pri + \delta_0^2} \bm{A}^\top (\bm{y} - \bm{A}\bm{x}_{A}^\pri)$
    \State Update $\bm{x}_{A}^\post$ by \eqref{lmmse}
    
    \State $v_{A}^\post = v_{A}^{pri} - \frac{M}{N} \frac{(v_{A}^{pri})^2}{v_{A}^{pri} + \delta^2}$ \label{eq:lmmse_va_post}
    \vspace{-2pt}
    \State $v_{A}^\ext = {\left( {1}/{v_{A}^\post} - {1}/{v_{A}^{pri}} \right)}^{-1}$ \label{eq:lmmse_va_ext}
    \vspace{-2pt}
    \State $\bm{x}_{A}^\ext = v_{A}^\ext \left({\bm{x}_{A}^\post}/{v_{A}^\post} - {\bm{x}_{A}^{pri}}/{v_{A}^{pri}} \right)$ \label{eq:lmmse_xa_ext}
    \vspace{-2pt}
    \State $\bm{x}_{B}^{pri} = \bm{x}_{A}^\ext,\; v_{B}^{pri} = v_{A}^\ext$ \label{eq:update_xb_pri}

    \State \text{\% Score-based MMSE denoiser}
    \State $\bm{x}_{B}^\post = \bm{x}_{B}^\pri + v_{B}^{pri} \bm{s}_{\bm{\theta}}(\bm{x}_{B}^{pri}, v_{B}^{pri})$ \label{eq:denoise_xb_post}
    % \State $v_{B}^\post = v_{B}^\pri + \frac{(v_{B}^\pri)^2}{N} \mathrm{tr}\left( \bm{S}_{\bm{\phi}}(\bm{x}_{B}^\pri, v_{B}^\pri) \right)$ \label{eq:denoise_vb_post}
    \State $v_B^{post} = \left({\| \bm{y} - \bm{A} \bm{x}_B^{post} \|_2^2 - M \delta^2}\right)/{\| \bm{A} \|_F^2}$ \label{eq:denoise_vb_post}
    \State $v_{B}^\ext = {\left( {1}/{v_{B}^\post} - {1}/{v_{B}^{pri}} \right)}^{-1}$ \label{eq:denoise_vb_ext}
    \vspace{-2pt}
    \State $\bm{x}_{B}^\ext = v_{B}^\ext \left({\bm{x}_{B}^\post}/{v_{B}^\post} - {\bm{x}_{B}^{pri}}/{v_{B}^{pri}} \right)$ \label{eq:denoise_xb_ext}
    \vspace{-2pt}
    % \State $\bm{x}_{B}^\ext = c \left( {\bm{x}_{B}^\post} - \alpha{\bm{x}_{B}^\pri} \right)$ \label{eq:denoise_xb_ext}
    \State $\bm{x}_{A}^{pri} = \bm{x}_{B}^\ext,\; v_{A}^{pri} = v_{B}^\ext$ \label{eq:update_xa_pri}
\Until{the stopping criterion is met}
\end{algorithmic}
\end{algorithm}

\subsection{Problem Description}
% The STMP algorithm exhibits certain limitations. Specifically, the effectiveness of the score-based MMSE denoiser relies on the input modeling assumption in \eqref{xb model}. Ideally, when the measurement matrix $\bm A$ satisfies the right-rotationally invariant property\cite{vamp} and the dimensions $M,\,N \xrightarrow{} \infty$ with a fixed ratio\cite{iidgaussian}, the elements of the error vector $\bm x_B^{pri}-\bm x$ is asymptotically i.i.d. Gaussian, thereby justifying \eqref{xb model}. However, in realistic settings, $\bm A$ may not possess the right-rotationally invariant property. Furthermore, the finite dimensions of $\bm A$ violate the asymptotic condition ($M,\,N \xrightarrow{} \infty$), causing the noise in \eqref{xb model} to deviate from the ideal i.i.d. Gaussian distribution. Consequently, a mismatch arises between the algorithm's underlying assumptions and the actual problem characteristics. To overcome these limitations, we adopt a data-driven strategy based on a deep unfolding framework. This approach empowers the network to adaptively learn and accommodate diverse noise distributions without relying on rigid Gaussian assumptions in \eqref{xb model}.
The STMP algorithm exhibits certain limitations. Specifically, the effectiveness of the score-based MMSE denoiser relies on the input modeling assumption in \eqref{xb model}. Ideally, when the measurement matrix $\bm A$ satisfies right-rotationally  invariant property and the denoiser is separable, the elements of the error vector $\bm x_B^{pri}-\bm x$ are asymptotically i.i.d. Gaussian as $M,\,N \xrightarrow{} \infty$ with a fixed ratio\cite{vamp}. However, in realistic settings, $\bm A$ is of finite dimensions and may not be right-rotationally invariant; the score-based network denoiser in STMP is non-separable. These factors, collectively, cause the error in \eqref{xb model} to deviate from the idealized i.i.d. Gaussian distribution. 
% Consequently, a mismatch arises between the algorithm's underlying assumptions and the actual problem characteristics. 
To overcome these limitations, we adopt a data-driven strategy based on a deep unfolding and refolding framework. This approach empowers the network to adaptively learn and accommodate diverse noise distributions without relying on the AWGN assumption in \eqref{xb model}.

\begin{figure*}
    \centering
    \includegraphics[width=1.0\linewidth]{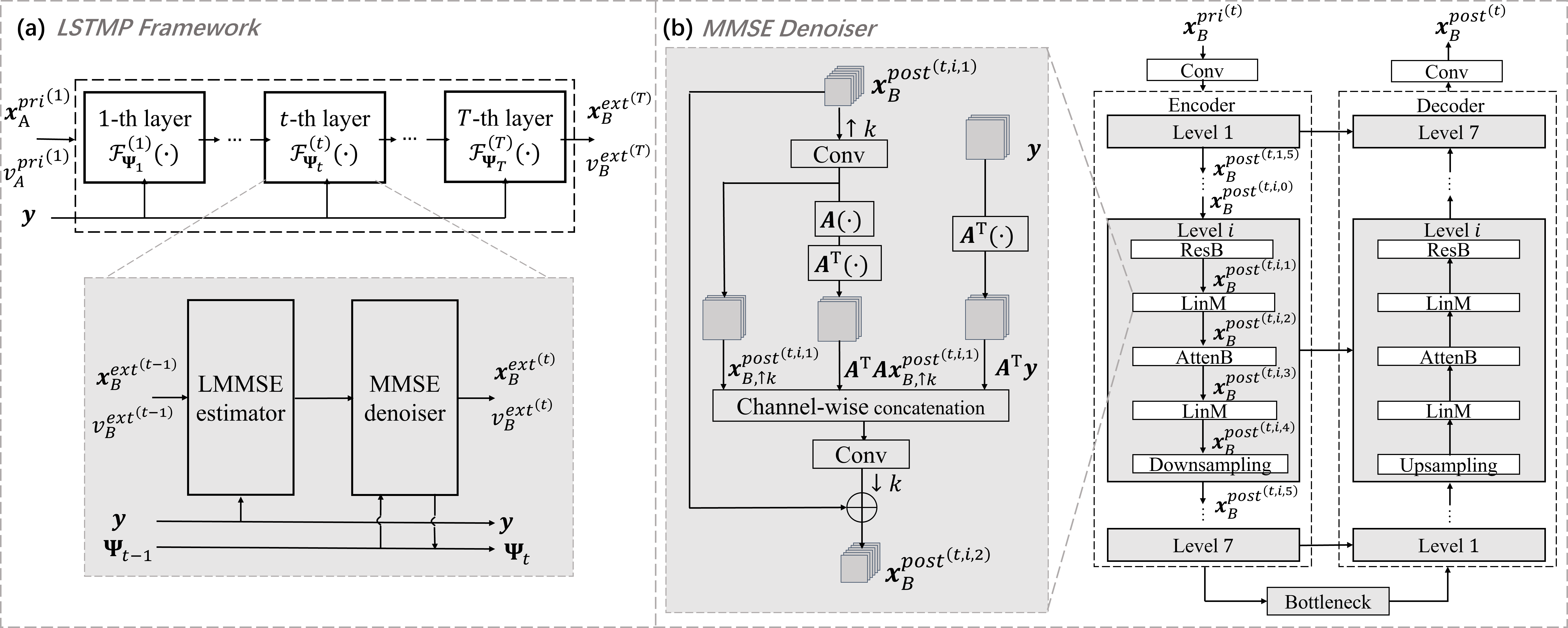}
    \caption{(a) The framework diagram of the LSTMP algorithm. (b) Refolding process within the noise estimation U-Net. ResB, AttenB and LinM denote the residual block, attention block and linear module, respectively. }
    % Both encoder and decoder have seven resolution levels $(1\leq i \leq 7)$, with linear modules in level $i$ inserted after each residual block and attention block when resolution level $i \leq 4$.}
    \label{fig:LSTMP framework}
\end{figure*}
\section{LSTMP Algorithm}
%\subsection{LSTMP Algorithm for partial orthogonal measurement matrix}
% In this section, we introduce the LSTMP algorithm to overcome the limitations of STMP mentioned in Section \ref{sec:section2}. LSTMP employs a data-driven strategy to fine-tune the score-based network at each layer, enabling the network to better handle noise other than AWGN.

% In this section, we introduce the LSTMP algorithm to overcome the limitations of STMP mentioned in Section \ref{sec:section2}. 
% LSTMP integrates a deep unfolding operation to unroll the iterative process and a refolding operation to treat the denoiser network as an iterative solver. This design effectively accelerates convergence and improves reconstruction quality.

\subsection{Deep Unfolding Framework}
\label{section_a}
Following the ideas of LAMP\cite{2017AMP} and LTMP\cite{2019Learned}, we construct a deep unfolding framework for STMP. Our deep unfolding framework consists of $T$ layers, where each layer corresponds to one iteration of STMP. In each layer, both the pre-trained network weights $\bm{\theta}$ and the iterative parameter $v_B^{pri}$ are fine-tuned. Accordingly, the parameter set for LSTMP is defined as $\bm\Psi=\{{v_B^{pri}}^{(t)},\,\bm\theta_t\}_{t=1}^T$, where ${v_B^{pri}}^{(t)}$ and $\bm\theta_t$ represent the parameters in the $t$-th layer. As shown in Fig. \ref{fig:LSTMP framework}(a), we unfold the iterative procedure of STMP into a $T$-layer feed-forward neural network: 
% $\mathcal{F}=\mathcal{F}_{\bm{\Psi}_1}^{(1)} \circ \mathcal{F}_{\bm{\Psi}_2}^{(2)} \circ \cdots  \mathcal{F}_{\bm{\Psi}_t}^{(t)} \circ \cdots \mathcal{F}_{\bm{\Psi}_T}^{(T)},$
\begin{equation}
    \mathcal{F}=\mathcal{F}_{\bm{\Psi}_1}^{(1)} \circ \mathcal{F}_{\bm{\Psi}_2}^{(2)} \circ \cdots  \mathcal{F}_{\bm{\Psi}_t}^{(t)} \circ \cdots \mathcal{F}_{\bm{\Psi}_T}^{(T)},
\end{equation}
where $\mathcal{F}_{\bm{\Psi}_t}^{(t)}$ represents the $t$-th layer of STMP parameterized with $\bm{\Psi}_t$, and $\circ$ refers to the connections between different layers. The update rule for the reconstructed result ${{\bm x}_B^{ext}}^{(t)}$ in the $t$-th layer is formulated as ${\bm{x}_B^{ext}}^{(t)}={\mathcal{F}^{(t)}_{\bm{\Psi}_t}}({\bm{x}_B^{ext}}^{(t-1)};\,\bm{y,\,\bm{A}}).$

\subsection{Refolded Denoiser as an Iterative Solver}

For the denoiser design, we adopt the U-Net\cite{unet} as the core architecture, given its superior performance in image reconstruction tasks. This architecture inherently establishes a symmetric compression (encoder) and decompression (decoder) structure. Both the encoder and decoder consist of multiple resolution levels, where residual blocks are employed to learn residual mappings and attention blocks are used to dynamically weight features. Essentially, U-Net is a deep neural network. By introducing a linear module for likelihood estimation after each residual and attention block (i.e. each layer) in U-Net, we treat each layer of the U-Net as an iteration between linear and non-linear modules.
% Leveraging the U-Net as a deep neural network, we formulate it as an iterative solver by introducing likelihood information across its resolution levels. By introducing a linear module after each residual and attention block in the noise estimation U-Net for likelihood estimation, we refold the denoiser network into an iterative solver that passes messages between linear and non-linear modules, where the non-linear operations are inherently provided by the residual and attention blocks. 
Since the data is compressed into feature maps with different resolutions at each layer of the U-Net, it cannot directly serve as a reconstructed image. Therefore, we need to upsample the feature maps back to the image domain for reconstruction. As illustrated in Fig. \ref{fig:LSTMP framework}(b), for a feature map ${\bm x_B^{post}}^{(t,i,j)}\in \mathbb{R}^{(H/k) \times (W/k) \times C}$, where $i$ corresponds to the current resolution level and $j$ indicates the specific processing stage within the level, the process of upsampling is given by
% By introducing a linear module after each residual and attention block in the noise estimation U-Net for likelihood estimation, we refold the denoiser network into an iterative solver that relies on turbo message passing between linear and non-linear modules, where the non-linear operations are inherently provided by the residual and attention blocks. This design enables the denoiser network to alternate between non-linear feature refinement and linear likelihood estimation, thereby satisfying the turbo message passing principle and effectively performing multiple turbo message passing iterations within the denoiser. Since the linear estimation is performed in the image domain, the feature map ${\bm x_B^{post}}^{(t,i,j)}\in \mathbb{R}^{(H/k) \times (W/k) \times C}$ in the network needs to be upsampled and transformed into the image domain, where $i$ corresponds to the current resolution level and $j$ indicates the specific processing stage within the level. Specifically, as illustrated in Fig. \ref{fig:LSTMP framework}(b), the process of transforming the feature map into the image domain is given by
\begin{equation}
    {\bm x_{B,\,\uparrow k}^{post}}^{(t,i,j)}=\text{Conv}\left(\left({\bm x_B^{post}}^{(t,i,j)}\right)_{\uparrow k}\right)\in \mathbb{R}^{H \times W \times 3},
    \label{upsample}
\end{equation}
where $\text{Conv}(\cdot)$ is a $3 \times3$ convolution and $\uparrow k / \downarrow k$ is the upsample / downsample operator by a factor of $k$. ${\bm x_{B,\,\uparrow k}^{post}}^{(t,i,j)}$ in \eqref{upsample} represents the upsampled feature map. The likelihood estimation shown in \eqref{lmmse} is essentially a linear combination of $\bm{x}_A^{pri}$, $\bm A^\top \bm y$ and $\bm A^\top \bm A \bm{x}_A^{pri}$. The incorporation of these physical measurement information into the refolding process is given by
\begin{equation}
\begin{split}
    & {\bm x_B^{post}}^{(t,i,j+1)} = {\bm x_B^{post}}^{(t,i,j)} \\
    & \quad +\left(\text{Conv}([{\bm x_{B,\uparrow k}^{post}}^{(t,i,j)}, \bm{A}^\top \bm{A} {\bm x_{B,\uparrow k}^{post}}^{(t,i,j)}, \bm{A}^\top \bm{y}])\right)_{\downarrow k},
    \label{downsample}
\end{split}
\end{equation}
% \begin{equation}
%      {\bm x_B^{post}}^{(t,i,j+1)} = {\bm x_B^{post}}^{(t,i,j)}
%      \quad +\text{Conv}(\bm z)_{\downarrow k},
%     \label{downsample}
% \end{equation}
where $[{\bm x_{B,\uparrow k}^{post}}^{(t,i,j)}, \bm{A}^\top \bm{A} {\bm x_{B,\uparrow k}^{post}}^{(t,i,j)}, \bm{A}^\top \bm{y}]$ refers to a channel-wise concatenation of  ${\bm x_{B,\,\uparrow k}^{post}}^{(t,i,j)}$, $\bm{A}^\top \bm{A} {\bm x_{B,\,\uparrow k}^{post}}^{(t,i,j)}$, and $\bm{A}^\top \bm{y}$. \eqref{downsample} introduces the likelihood estimation and transforms the image domain data ${\bm x_{B, \uparrow k}^{post}}^{(t,i,j)}$ back to the feature map ${\bm x_B^{post}}^{(t,i,j+1)}$.

While deep unfolding unfolds an iterative solver into a deep neural network, \eqref{upsample} and \eqref{downsample} reverse this process by converting a deep neural network into an iterative solver. Therefore, we refer to this procedure as "refolding".

\subsection{Training Strategy}

% For network training, we employ normalized mean square error (NMSE)
% % \begin{equation}
% %     f(\bm\Psi)=\frac{\lVert \bm{x}_B^{ext}-\bm{x} \rVert^2}{\lVert \bm{x}\rVert^2} 
% % \end{equation}
% as the loss function. To improve training stability and avoid overfitting, the training procedure follows a layer-by-layer training method\cite{2017AMP}. 
% Specifically, the optimized parameter set $\bm\Psi_{t-1}$ from the $(t-1)$-th layer is used to initialize the parameter set $\bm\Psi_{t}$ in the $t$-th layer. 
% Parameters $\bm \Psi_t$ in the $t$-th layer are optimized by backpropagation to minimize the loss function
% \begin{equation}
%     f^{(t)}(\bm\Psi_t)=\frac{\lVert {{\bm{x}_B^{ext}}^{(t)}-\bm{x}} \rVert^2}{\lVert \bm{x}\rVert^2}.
% \end{equation} 

% For the network parameters $\bm \theta$, we introduce three initialization methods:
% (1) Initialize $\bm \theta$ randomly; (2) use the pre-trained score-based network parameters from STMP; (3) use a hybrid training strategy. In hybrid training, for the first layer $\mathcal{F}_{\bm{\Psi_1}}^{(1)}$ of LSTMP, we initialize $\bm \theta$ randomly; for the second layer $\mathcal{F}_{\bm{\Psi_{2}}}^{(t)}$, we use the pre-trained score-based network parameters from STMP. 

For network training, we adopt a layer-by-layer training strategy\cite{2019Learned}, where the optimized parameters $\bm\Psi_{t-1}$ from the $(t-1)$-th layer are used to initialize the parameters $\bm\Psi_{t}$ in the $t$-th layer. At the $t$-th layer, the parameters are optimized via backpropagation using the normalized mean square error (NMSE) as the loss function:
\begin{equation}
    f^{(t)}(\bm\Psi_t)=\frac{\lVert {{\bm{x}_B^{ext}}^{(t)}-\bm{x}} \rVert^2}{\lVert \bm{x}\rVert^2}.
\end{equation} 
For the network weights $\{\bm\theta_t\}_{t=1}^T$, we consider three initialization strategies: (1)  Initialize $\bm \theta_1$ in the first layer randomly; (2) use the pre-trained score-based network weights from STMP to initialize $\bm \theta_1$; (3) use a hybrid initialization strategy, where $\bm \theta_1$ in the first layer is randomly initialized and $\bm \theta_2$ in the second layer uses the score-based network weights from STMP.

\section{Experiments}
\label{sec:section5}
\subsection{Experimental Settings}

% In this section, we evaluate the results of the LSTMP on the FFHQ dataset\cite{2019A}. The baseline algroithms include STMP\cite{2025Score}, DMPS\cite{dmps}, DiffPIR\cite{diffpir}, DDRM\cite{ddrm} and IDM\cite{IDM}. We evaluate the algorithm performance on a compressive image recovery task.
In this section, we evaluate LSTMP on the FFHQ dataset \cite{2019A} for compressive image recovery, comparing it with STMP \cite{2025Score}, DMPS \cite{dmps}, DiffPIR \cite{diffpir}, DDRM \cite{ddrm}, and IDM \cite{IDM}. 
The measurement matrix is chosen as $\bm{A=SW\!D}$\cite{XJAZ03B41BBF891512CD912CDF1605452A7F}, where $\bm{S} \in \mathbb{R}^{M \times N}$ is a random row-selection matrix, $\bm{W} \in \mathbb{R}^{N \times N}$ is a DCT matrix and $\bm{D} \in \mathbb{R}^{N \times N}$ is a diagonal matrix with $1$s and $-1$s on the diagonal. Experiments are conducted under a compression ratio $M/N=0.1$ with noise levels $\delta=0.05$ and $\delta=0.5$. 
LSTMP terminates when the relative change of $\bm x_B^{ext}$ between two consecutive iterations falls below $6 \times 10^{-3}$. 

\subsection{Ablation Study}
As shown in Table \ref{tab:ablation study}, we conduct an ablation study at a noise level of $\delta=0.05$, with NMSE (dB) at layers 1, 3, and 5 as the evaluation metric. "Init.", "Rand.", "Pret.", and "Hyb." denote initialization, random initialization, pre-trained weights from STMP, and hybrid initialization, respectively. STMP-1, STMP-2, and LSTMP-1 to LSTMP-5 denote different configurations in the ablation study.
% LSTMP-1 $vs.$ LSTMP-4, LSTMP-2 $vs.$ LSTMP-5 and LSTMP-3 $vs.$ LSTMP (proposed) show that refolding can effectively improve reconstruction quality. 
We can observe that the results obtained using unfolding significantly outperform those without unfolding. 
The comparisons of STMP-1 with STMP-2, LSTMP-1 with LSTMP-4, LSTMP-2 with LSTMP-5, and LSTMP-3 with the proposed LSTMP demonstrate the effectiveness of refolding.
% By comparing (1) with (2), (3) with (6), (4) with (7), and (5) with the proposed (8) under the same settings except for the refolding strategy, 
% LSTMP-1 $vs.$ LSTMP-4, LSTMP-2 $vs.$ LSTMP-5 and LSTMP-3 $vs.$ LSTMP (proposed) show that refolding can effectively improve reconstruction quality.
For initialization strategies, random initialization outperforms using pre-trained weights at layer 1 but yields inferior results at layers 3 and 5. We conjecture that this is because the pretrained weights rely on the i.i.d. Gaussian assumption, while the error in \eqref{xb model} deviates from such a distribution in the initial iterations. 
Therefore, we adopt a hybrid initialization strategy that employs random initialization for layer 1 and uses pretrained weights for layer 2. This strategy improves reconstruction quality across all three layers. Ultimately, the proposed method, which leverages both refolding and hybrid training, achieves the best overall performance among all the configurations.

\subsection{Comparison with Baseline Algorithms}
\begin{figure}[t]
    \raggedright
    \setlength{\tabcolsep}{1pt} % 减少列间距
    
    \begin{tabular}{c c}
        \rotatebox[origin=c]{90}{\footnotesize $\delta = 0.05$} &
        \begin{minipage}{0.90\linewidth}
            \centering
            \begin{tabular}{c c c c c c}
            
                \includegraphics[width=0.165\linewidth]{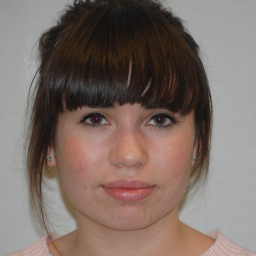} &
                \includegraphics[width=0.165\linewidth]{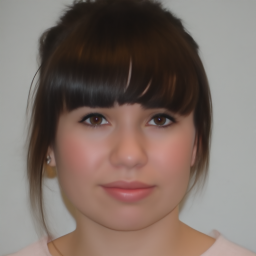} &
                \includegraphics[width=0.165\linewidth]{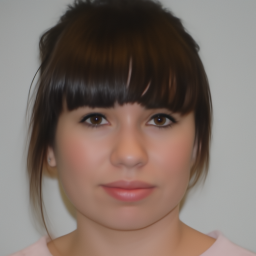} &
                \includegraphics[width=0.165\linewidth]{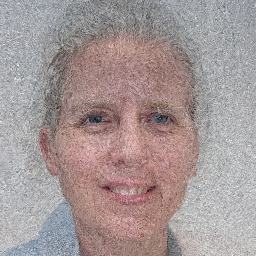} &
                \includegraphics[width=0.165\linewidth]{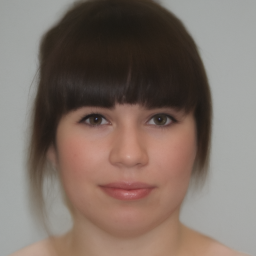} &
                \includegraphics[width=0.165\linewidth]{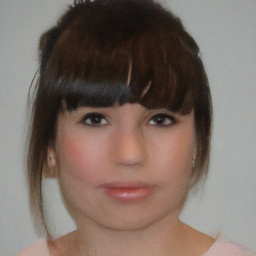} \\[-3pt]
                
                \includegraphics[width=0.165\linewidth]{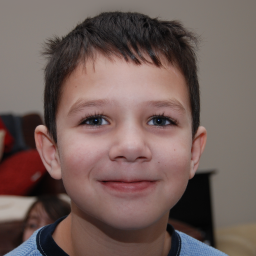} &
                \includegraphics[width=0.165\linewidth]{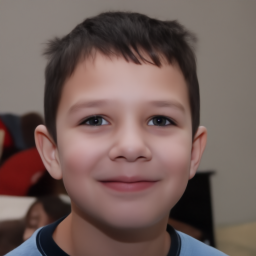} &
                \includegraphics[width=0.165\linewidth]{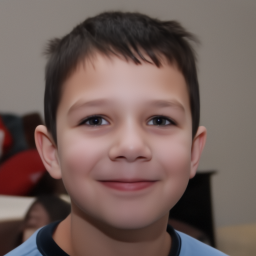} &
                \includegraphics[width=0.165\linewidth]{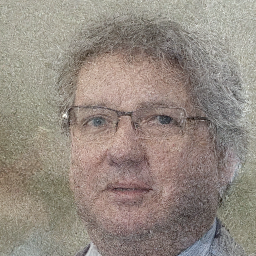} &
                \includegraphics[width=0.165\linewidth]{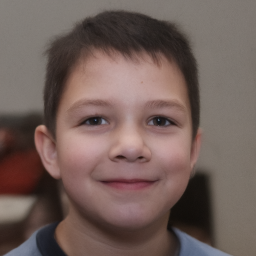} &
                \includegraphics[width=0.165\linewidth]{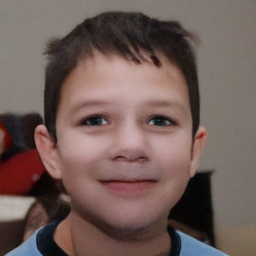} \\[-3pt]
            \end{tabular}
        \end{minipage} \\
        
        \rotatebox[origin=c]{90}{\footnotesize $\delta = 0.5$} &
        \begin{minipage}{0.90\linewidth}
            \centering
            \begin{tabular}{c c c c c c}
                \includegraphics[width=0.165\linewidth]{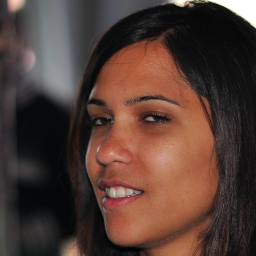} &
                \includegraphics[width=0.165\linewidth]{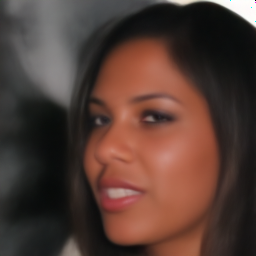} &
                \includegraphics[width=0.165\linewidth]{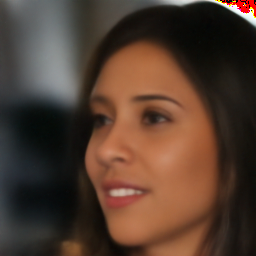} &
                \includegraphics[width=0.165\linewidth]{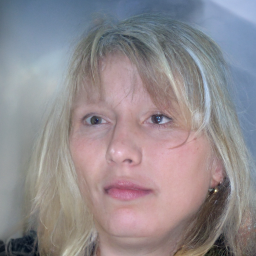} &
                \includegraphics[width=0.165\linewidth]{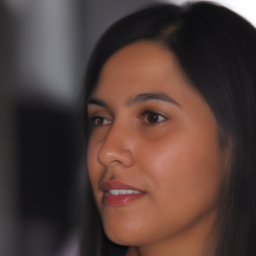} &
                \includegraphics[width=0.165\linewidth]{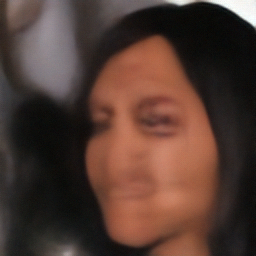} \\[-3pt]
                
                \includegraphics[width=0.165\linewidth]{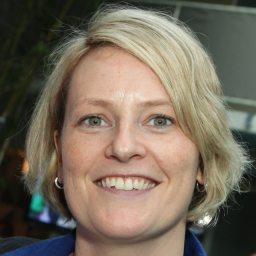} &
                \includegraphics[width=0.165\linewidth]{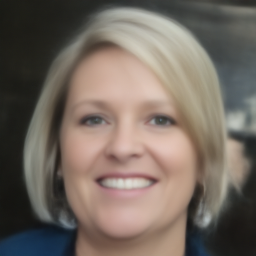} &
                \includegraphics[width=0.165\linewidth]{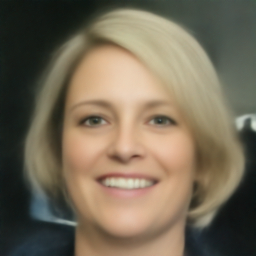} &
                \includegraphics[width=0.165\linewidth]{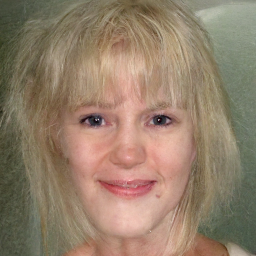} &
                \includegraphics[width=0.165\linewidth]{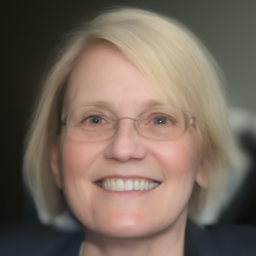} &
                \includegraphics[width=0.165\linewidth]{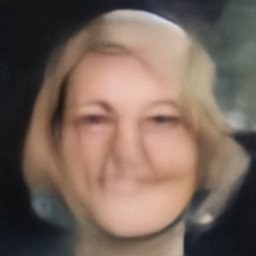} \\[-3pt]
                \scriptsize Ground truth & \scriptsize LSTMP & \scriptsize STMP & \scriptsize DMPS & \scriptsize DDRM & \scriptsize IDM \\
            % \end{tabular}
            % \vspace{2mm}
            % \begin{tabular}{c c c c c}
                % \scriptsize (a) Ground truth &
                % \scriptsize (b) LSTMP-HI (Ours) &
                % \scriptsize (c) STMP &
                % % \scriptsize (d) DMPS &
                % % \scriptsize (e) DiffPIR &
                % \scriptsize (f) DDRM &
                % \scriptsize (g) IDM \\
            \end{tabular}
        \end{minipage}
    \end{tabular}
    \medskip
    \raggedright
    \scriptsize

    % \hspace{0.5em} Ground truth \hspace{0.1em}
    % LSTMP-HI \hspace{1.5em}
    % STMP \hspace{1.9em}
    % DMPS \hspace{1.8em}
    % DDRM \hspace{2.3em}
    % IDM
    \vspace{-10pt} 
    \caption{\footnotesize Partial results of compressed image reconstruction on the FFHQ dataset with compression ratio $M/N=0.1$.}
    \label{fig:cr_image}
\end{figure}

\begin{figure}[t]
    \centering
    % trim=左 下 右 上 (单位通常是pt或mm)，clip表示真正裁剪
    \includegraphics[width=1.0\linewidth, trim=0mm 7mm 0mm 1mm, clip]{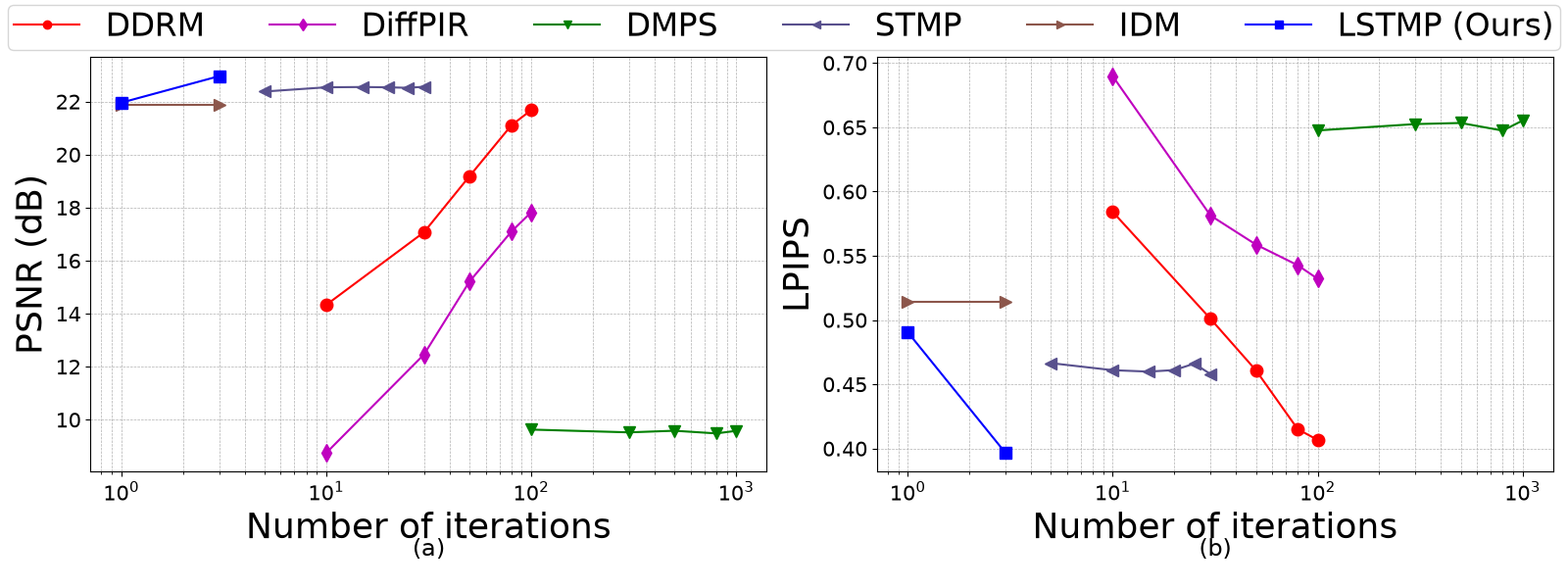}
    \vspace{-20pt} 
    \caption{The tradeoff between algorithm complexity and reconstruction quality. The noise level is $\delta=0.5$.}
    \label{fig:complexity}
\end{figure}

\begin{table}[t]
    \centering
    \caption{Ablation study of LSTMP. "Rand.", "Pret.", and "Hyb." denote random, pretrained, and hybrid initialization, respectively.}
    \begin{tabular}{ l @{} >{\centering\arraybackslash}p{0.9cm} >{\centering\arraybackslash}p{0.8cm} >{\centering\arraybackslash}p{0.6cm} c }
        \toprule
        \textbf{Method} & Unfold & Init. & Refold & NMSE\,(dB) for layer 1\,/\,3\,/\,5 \\
        \midrule  % ← 只用这一条线！
        % LSTMP-1 & Rand. & \ding{55} & -21.65 / 24.61 \\
        % LSTMP-2 & Pret. & \ding{55} & -17.81 / -24.38\\
        % LSTMP-3 & Hyb. & \ding{55} &   /   \\
        % LSTMP-4 & Rand. & \ding{51} & -24.53 / -24.69  \\
        % LSTMP-5 & Pret. & \ding{51} & -21.84 / -24.80 \\
        % LSTMP (Proposed) & Hyb. & \ding{51} & \textbf{-24.54} / \textbf{-25.05} \\
        STMP-1  & \ding{55} & -     & \ding{55} & $-$4.12\,/\,$-$0.28\,/\,$-$3.23 \\
        STMP-2  & \ding{55} & -     & \ding{51} & $-$8.60\,/\,$-$7.53\,/\,$-$11.30 \\
        LSTMP-1 & \ding{51} & Rand. & \ding{55} & $-$21.37\,/\,$-$24.26\,/\,$-$24.54 \\
        LSTMP-2 & \ding{51} & Pret. & \ding{55} & $-$17.81\,/\,$-$24.38\,/\,$-$25.01 \\
        LSTMP-3 & \ding{51} & Hyb. & \ding{55}  & $-$21.41\,/\,$-$24.54\,/\,$-$25.10 \\
        LSTMP-4 & \ding{51} & Rand. & \ding{51} & $-$24.53\,/\,$-$24.69\,/\,$-$24.76 \\
        LSTMP-5 & \ding{51} & Pret. & \ding{51} & $-$21.84\,/\,$-$24.80\,/\,$-$25.23 \\
        LSTMP (Ours)  & \ding{51} & Hyb. & \ding{51} & \textbf{$-$24.54}\,/\,\textbf{$-$25.05} / \textbf{$-$25.25} \\
        \bottomrule
    \end{tabular}
    \label{tab:ablation study}
\end{table}

\begin{table}[t]
    \centering
    % \caption{Quantitative comparisons of performance on FFHQ dataset at noise levels $\delta=0.05\,/\,\delta=0.5$.}
    \caption{Quantitative comparisons at two noise levels. Values before and after "/" correspond to $\delta=0.05$ and $\delta=0.5$, respectively.}
    \label{tab:comparison}
    % 修正点：将 {lllcc} 改为 {llcccc}，确保有6列来容纳所有数据
    \begin{tabular}{@{} l @{} cccc @{}} 
        \toprule
        % 表头部分：不需要 multirow，直接列出即可
         {Algorithm} & {NFEs} $\downarrow$ & {SSIM} $\uparrow$ & {PSNR (dB)} $\uparrow$ & {LPIPS} $\downarrow$ \\
        
        \midrule
        % 数据行
        % 注意：第一列留空表示 Task 分组（如果需要显示 Task 名称，可以在第一行填入）
          LSTMP (Ours)  
        & \textbf{3} & \textbf{{0.8695}\,/\,{0.6634}} & \textbf{{30.89}\,/\,\textbf{22.96}} & \textbf{0.2064}\,/\,\textbf{0.3966} \\
         
          STMP 
        & 30  & 0.8635\,/\,0.6407 & 30.53\,/\,22.56 & 0.2198\,/\,0.4580 \\

          DiffPIR  
        & 100 & 0.6120\,/\,0.5201 & 21.13\,/\,17.82 & 0.4676\,/\,0.5321 \\
   
          DMPS  
        & 1000& 0.1035\,/\,0.2200 & 10.88\,/\,9.580 & 0.7517\,/\,0.6553 \\
      
          DDRM  
        & 100 & 0.7147\,/\,0.6293 & 24.63\,/\,21.68 & 0.3162\,/\,0.4067 \\

          IDM
        & \textbf{3}   & 0.8005\,/\,0.5941 & 29.37\,/\,21.88 & 0.2563\,/\,0.5144 \\
        
        \bottomrule
    \end{tabular}
\end{table}

Fig. \ref{fig:cr_image} presents some results of compressed image reconstruction. DMPS fails to recover the original images at both noise levels. At a low noise level, LSTMP, STMP, DDRM and IDM all present visually high-quality reconstructions. At a high noise level, IDM generates nearly invalid images, indicating its poor performance under high noise. DDRM outperforms IDM but occasionally generates non-existent features in the original image.
% For example, in the last row of Fig. \ref{fig:cr_image}, DDRM generates a woman wearing glasses, while the ground truth shows a woman without glasses. 
While both LSTMP and STMP yield results relatively consistent with the ground truth, LSTMP preserves structural details more clearly in the presented examples. These results demonstrate that LSTMP achieves favorable reconstruction fidelity across different noise levels.

Fig. \ref{fig:complexity} illustrates the tradeoff between computational complexity and reconstruction quality, evaluated via peak signal-to-noise ratio (PSNR) and learned perceptual image patch similarity (LPIPS)\cite{lpips}. The computational complexity is dominated by the number of neural function evaluations (NFEs). 
Since all methods require 1 NFE per iteration, the number of iterations represents their computational cost. 
DDRM, DiffPIR, and STMP suffer from high computational complexity. DMPS generates invalid images. IDM has a comparable complexity to LSTMP, but its reconstruction performance is significantly inferior.
LSTMP surpasses most baselines in both PSNR and LPIPS with just one iteration and outperforms all baselines within 3 iterations, demonstrating its effectiveness in balancing reconstruction quality and computational cost.
% These results suggest that LSTMP achieves a favorable trade-off between computational cost and reconstruction quality, highlighting the effectiveness of the proposed network design.}

Table \ref{tab:comparison} compares LSTMP with other baseline algorithms in terms of NFEs, structural similarity index measure (SSIM)\cite{ssim}, PSNR and LPIPS. It is worth noting that LSTMP achieves the best performance across all metrics with only 3 NFEs, which is highly advantageous compared to most of the baseline algorithms. These results indicate that our method not only achieves pixel-level accuracy but also recovers more perceptually consistent images, while significantly reducing computational complexity.

\section{Conclusion} 
In this letter, we introduced LSTMP for compressive image recovery. LSTMP employs a deep unfolding and refolding framework, enabling the algorithm to alternate between an unrolled network and an iterative solver, achieving fast convergence while maintaining high reconstruction quality. Extensive experiments on the FFHQ dataset demonstrated its effectiveness against various state-of-the-art baseline algorithms. Notably, LSTMP achieves this superior performance with only three iterations. 
% Furthermore, LSTMP can be extended beyond compressive image recovery to channel recovery, source separation, and other linear inverse problems.
Furthermore, the proposed LSTMP framework is not restricted to image recovery, but has the potential to be extended to a wide range of applications, including channel recovery, source separation, and other linear inverse problems.

% You can use a bibliography generated by BibTeX as a .bbl file.
%  BibTeX documentation can be easily obtained at:
%  http://mirror.ctan.org/biblio/bibtex/contrib/doc/
%  The IEEEtran BibTeX style support page is:
%  http://www.michaelshell.org/tex/ieeetran/bibtex/
 

\IEEEtriggeratref{15}
\bibliographystyle{IEEEtran}
\bibliography{main}

\clearpage
\section{Supplementary Material}
\subsection{Additional Experiments under Different Settings}
To evaluate LSTMP beyond the main experimental setting, we consider an alternative measurement operator, a higher compression ratio, and an additional dataset. All experiments in this subsection use a noise level of $\delta=0.05$.

    \begin{table}[H]
        \centering
        \caption{Comparison results between LSTMP and baseline methods under a sampling ratio of 0.1, a noise level of 0.05, and a structurally random, orthonormalized i.i.d. Gaussian measurement matrix $\bm A$.}
        \label{tab:metric_comparison1}
        \begin{tabular}{ccccc}
        \toprule
        Method & NFEs & PSNR (dB) & SSIM & LPIPS \\
        \midrule
        LSTMP & \textbf{$\bm {3}$} & \textbf{$\bm {30.88}$} & \textbf{$\bm {0.8691}$} & \textbf{$\bm {0.2067}$} \\
        STMP  & $30$ & $30.57$ & $0.8637$ & $0.2202$ \\
        DDRM  & $100$ & $24.71$ & $0.7168$ & $0.3138$ \\
        \bottomrule
        \end{tabular}
    \end{table}
    
    Table \ref{tab:metric_comparison1} reports results with a structurally random, orthonormalized i.i.d. block-based Gaussian measurement matrix \cite{IDM}, where the block size is fixed at $32 \times 32$. This provides an additional operator setting beyond the original $\bm{A} = \bm{SW\!D}$ configuration, allowing us to evaluate the generalizability of LSTMP under different measurement structures. 
    % The compression ratio is 0.1 and the noise level is 0.05. As can be observed, LSTMP consistently outperforms STMP and DDRM in terms of PSNR, SSIM, and LPIPS, which indicates that LSTMP achieves improvements not only in pixel-wise reconstruction fidelity but also in structural similarity and perceptual quality.

    \begin{table}[H]
        \centering
        \caption{Results on the FFHQ dataset with a sampling rate of 0.2 and a noise level of 0.05.}
        \label{tab:metric_comparison2}
        \begin{tabular}{ccccc}
        \toprule
        Method & NFEs & PSNR (dB) & SSIM & LPIPS \\
        \midrule
        LSTMP & \textbf{$\bm 3$} & \textbf{$\bm {32.71}$} & \textbf{$\bm {0.9020}$} & \textbf{$\bm {0.1696}$} \\
        STMP  & $30$ & $32.52$ & $0.9012$ & $0.1887$ \\
        DDRM  & $100$ & $29.47$ & $0.8381$ & $0.2080$\\
        \bottomrule
        \end{tabular}
    \end{table}

    Table~\ref{tab:metric_comparison2} examines a higher compression ratio of $M/N=0.2$ on FFHQ, using the original $\bm{A}=\bm{SW\!D}$ operator. This complements the original $M/N = 0.1$ setting. 
    % Table \ref{tab:metric_comparison2} presents results on the FFHQ dataset with a compression ratio of $M/N = 0.2$ and a noise level of $\delta=0.05$. This complements the original $M/N = 0.1$ setting. 
    % As shown in the table, LSTMP achieves the best performance across all three reconstruction metrics, with a PSNR of 32.71 dB, an SSIM of 0.9020, and an LPIPS of 0.1696. Compared with STMP, LSTMP provides improvements in both reconstruction fidelity and perceptual quality while requiring only 3 NFEs, compared with 30 NFEs for STMP and 100 NFEs for DDRM. These results further demonstrate the effectiveness and computational efficiency of LSTMP under a different compression ratio.  

    \begin{table}[H]
            \centering
            \caption{Results on the Lsun bedroom dataset with a sampling rate of 0.2 and a noise level of 0.05.}
            \label{tab:metric_comparison3}
            \begin{tabular}{ccccc}
            \toprule
            Method & NFEs & PSNR (dB) & SSIM & LPIPS \\
            \midrule
            LSTMP & \textbf{$\bm 3$} & \textbf{$\bm{31.56}$} & \textbf{$\bm{0.8975}$} & \textbf{$\bm{0.1668}$} \\
            STMP  & $30$ & $30.98$ & $0.8819$ & $0.1892$ \\
            DDRM  & $100$ & $27.26$ & $0.7744$ & $0.2807$\\
            \bottomrule
            \end{tabular}
    \end{table}
    Table \ref{tab:metric_comparison3} further evaluates LSTMP on LSUN Bedroom \cite{lsun}, using the measurement operator $\bm{A}=\bm{SW\!D}$ and the compression ratio $M/N=0.2$. It provides an additional natural-image benchmark consistent with the paper's motivation.
    % To improve dataset diversity, we have added experiments on the LSUN bedroom dataset \cite{lsun} under $M/N = 0.2$ and noise level $0.05$. The results are shown in Table \ref{tab:metric_comparison3} and provide an additional natural-image benchmark consistent with the paper's motivation.

    Across these settings, LSTMP achieves the highest PSNR and SSIM and the lowest LPIPS among the compared methods, using 3 NFEs versus 30 for STMP and 100 for DDRM. These results support the applicability of the proposed framework beyond the original dataset, measurement operator, and compression ratio.

\subsection{Statistical Analysis of LSTMP and STMP}
\begin{table}[t]
\centering
\caption{Comparison of average PSNR (dB, $\uparrow$), SSIM ($\uparrow$) and LPIPS ($\downarrow$)(±std) on FFHQ dataset with a compression ratio of 0.1 and a noise level of 0.05.}
\label{tab:std}
\begin{tabular}{lccc}
\toprule
Method & PSNR (dB) & SSIM & LPIPS \\
\midrule
LSTMP
& \textbf{$\bm {30.89} \pm \bm{1.7418}$}
& \textbf{$\bm{0.8695} \pm \bm{0.0375}$}
& \textbf{$\bm{0.2064} \pm \bm{0.0322}$} \\
STMP
& {$30.53 \pm 1.9480$}
& $0.8635 \pm 0.0413$
& $0.2198 \pm 0.0346$ \\
\bottomrule
\end{tabular}
\end{table}

\begin{table}[t]
\centering
\caption{Significance analysis of reconstruction performance.}
\label{tab:significance}
\begin{tabular}{lc}
\hline
Metric & $p$-value \\
\hline
PSNR       & $2.542 \times 10^{-11}$  \\
SSIM       & $1.781 \times 10^{-6}$   \\
LPIPS  & $7.406 \times 10^{-24}$  \\
\hline
\end{tabular}
\end{table}

To further assess the variability of the reconstruction performance, we have calculated the mean and standard deviation of PSNR, SSIM and LPIPS across the 100 test images. We provide the detailed mean $\pm$ standard deviation results in Table \ref{tab:std}. As shown, LSTMP achieves the better average reconstruction performance than STMP. 
  
The corresponding significance test is shown in Table \ref{tab:significance}. We employ a two-sided paired Student's t-test to compare LSTMP and STMP across the test images. With a significance level of 0.05, $p$-values below this threshold indicate statistically significant differences between STMP and LSTMP, leading to the rejection of the null hypothesis that the two methods exhibit no systematic performance difference. Indeed, all three metrics yield $p$-values well below 0.05, providing strong statistical evidence that the performance improvements of LSTMP over STMP are unlikely to be attributable to random variation.

\subsection{Implementation Details}

We use 30000 images from the FFHQ dataset for training and another 100 images for testing. The first and second layers are trained for 100 epochs, and each subsequent layer for 25 epochs, with 600 parameter updates per epoch. The batch size is set to 4. The Adam optimizer is adopted with an initial learning rate of $0.0001$, which is halved at each quarter of the training epochs. All experiments are conducted on a Linux server equipped with Intel Xeon Platinum 8358P CPUs and NVIDIA A800 GPUs. The prior variance $v_{B}^{pri}$ is parameterized as a learnable parameter and initialized to 1.0. 
% For different noise levels, i.e., $\delta=0.05$ and $\delta=0.5$, separate networks are trained independently. 
When comparing with the baseline algorithms, the unrolled depth \(T\), which corresponds to the number of iterations of the proposed algorithm, is set to 3. In the ablation study, the maximum value of \(T\) is set to 5. During testing, the measurement matrix $\bm{A}$ is independently generated for each test image, while the same measurement matrix is used for all compared methods on the same image.

\subsection{Additional Results on Ablation Study}
\begin{table*}[t]
\centering
\caption{Ablation study of different LSTMP configurations. The three values separated by "/" correspond to the results at layer 1, layers 3, and layers 5, respectively.}
\label{tab:ablation_config}
\footnotesize
\begin{tabular}{ccccccc}
\toprule
Method & Unfold & Init. & Refold & NMSE (dB)$\downarrow$ & PSNR (dB)$\uparrow$ & SSIM $\uparrow$  \\
\midrule
STMP-1  & \ding{55} & - & \ding{55} & $-4.12$ / $-0.28$ / $-3.23$ & $9.97$ / $6.13$ / $9.08$ & $0.0522$ / $0.0058$ / $0.0297$ \\
STMP-2  & \ding{55} & - & \ding{51} & $-8.60$ / $-7.53$ / $-11.30$ & $14.45$ / $13.38$ / $17.15$ & $0.1186$ / $0.1592$ / $0.2443$ \\
LSTMP-1 & \ding{51} & Rand. & \ding{55} & $-21.37$ / $-24.26$ / $-24.54$ & $27.23$ / $30.11$ / $30.39$ & $0.7756$ / $0.8511$ / $0.8573$ \\
LSTMP-2 & \ding{51} & Pret. & \ding{55} & $-17.81$ / $-24.38$ / $-25.01$ & $23.67$ / $30.23$ / $30.87$ & $0.6404$ / $0.8430$ / $0.8693$  \\
LSTMP-3 & \ding{51} & Hyb. & \ding{55} & $-21.41$ / $-24.54$ / $-25.10$ & $27.26$ / $30.39$ / $30.95$ & $0.7788$ / $0.8587$ / $0.8719$ \\
LSTMP-4 & \ding{51} & Rand. & \ding{51} & $-24.53$ / $-24.69$ / $-24.76$ & $\bm{30.38}$ / $30.54$ / $30.61$ & $0.8589$ / $0.8612$ / $0.8632$ \\
LSTMP-5 & \ding{51} & Pret. & \ding{51} & $-21.84$ / $-24.80$ / $-25.23$ & $27.68$ / $30.66$ / $31.08$ & $0.7693$ / $0.8653$ / $0.8730$ \\
LSTMP (Ours) & \ding{51} & Hyb. & \ding{51} & \textbf{$\bm {-24.54}$} / \textbf{$\bm {-25.05}$} / \textbf{$\bm {-25.25}$} & \textbf{$\bm {30.38}$} / \textbf{$\bm {30.89}$} / \textbf{$\bm {31.10}$} & \textbf{$\bm {0.8600}$} / \textbf{$\bm {0.8695}$} / \textbf{$\bm {0.8735}$} \\
\bottomrule
\end{tabular}
\end{table*}
We provide the corresponding PSNR and SSIM results of the ablation study in Table \ref{tab:ablation_config}. It can be seen that LSTMP (Ours) achieves the best performance.

\subsection{Additional Results on Complexity Analysis}
\begin{table}[t]
\centering
\caption{Comparison of per-layer and overall computational complexity and reconstruction performance.}
\setlength{\tabcolsep}{3pt}
\label{tab:per layer and total}
\begin{tabular}{llcccc}
\toprule
 & Method & NFEs & Inference time (s) & FLOPs (G) & PSNR (dB) \\
\midrule
\multirow{3}{*}{Per layer}
& LSTMP & $1$ & $0.142$ & $275.66$ & $\bm{30.38}$ \\
& STMP   & $1$ & $0.092$ & $266.74$ & $9.97$ \\
& DDRM   & $1$ & $\bm{0.031}$ & $\bm{194.01}$ & $6.98$ \\
\midrule
\multirow{3}{*}{Overall}
& LSTMP & $\bm{3}$ & $\bm{0.440}$ & $\bm{826.98}$ & $\bm{30.89}$ \\
& STMP   & $30$ & $2.464$ & $8002.14$ & $30.53$ \\
& DDRM   & $10$0 & $2.856$ & $19401.06$ & $24.63$ \\
\bottomrule
\end{tabular}
\end{table}

We have provided an explicit complexity analysis, including the cost per layer and the overall cost, in Table \ref{tab:per layer and total}. We use the inference time, FLOPs and PSNR to evaluate the trade-off between algorithm complexity and reconstruction quality. For STMP, "per layer" refers to per iteration, while for DDRM, it refers to one reverse diffusion step. In terms of computational complexity, LSTMP indeed requires the longest inference time and the highest FLOPs per layer. However, since it only requires three layers (i.e., 3 NFEs), it achieves significant advantages in terms of overall inference time and FLOPs. Regarding reconstruction quality, LSTMP achieves the best performance in both per-layer and overall comparisons.

\subsection{Convergence Analysis}

\begin{figure}
    \centering
    \includegraphics[width=1.0\linewidth]{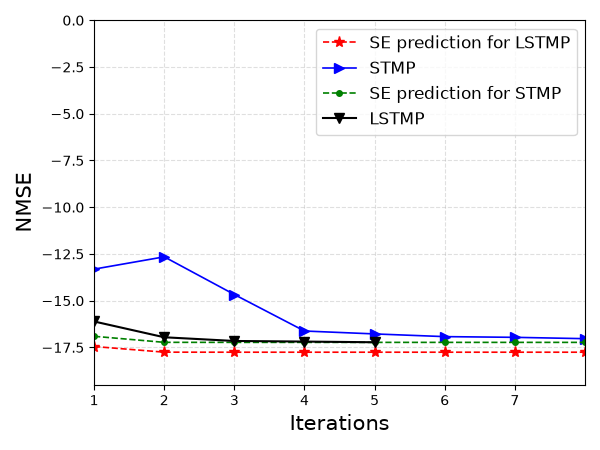}
    \caption{The NMSE trajectories
of STMP and LSTMP, along with their respective state evolution (SE) curves over iterations.}
    \label{fig:se}
\end{figure}

Fig. \ref{fig:se} shows the NMSE trajectories of STMP and LSTMP, along with their respective state evolution (SE) curves over iterations on the FFHQ dataset, under a compression rate of 0.1 and a noise level of 0.5. It can be observed that the SE curves of LSTMP and STMP are relatively close. However, LSTMP approaches convergence in just 3 iterations, whereas STMP requires nearly 6 iterations. This demonstrates that LSTMP achieves faster convergence.

\subsection{the Mismatch with the AWGN Assumption}

\begin{table}[t]
    \centering
    \caption{$p$-values of the D'Agostino--Pearson normality test for the error $\bm x_B^{pri}-\bm x$. The $p$-values below 0.05 are highlighted in bold.}
    \label{tab:normality_test_822}
    \begin{tabular}{c|cccc}
        \toprule
        Channel & $\text{Iter. }2$ & $\text{Iter. }3$ & $\text{Iter. }4$ & $\text{Iter. }5$ \\
        \midrule
        $\mathrm{R}$
        & $0.9320$
        & $0.0690$
        & $\mathbf{4.92\times10^{-4}}$
        & $\mathbf{5.57\times10^{-4}}$ \\

        $\mathrm{G}$
        & $\mathbf{0.0103}$
        & $\mathbf{0.0125}$
        & $0.1472$
        & $0.7425$ \\

        $\mathrm{B}$
        & $0.8421$
        & $0.0738$
        & $\mathbf{0.0020}$
        & $\mathbf{0.0107}$ \\
        \bottomrule
    \end{tabular}
\end{table}

\begin{figure*}
    \centering
    \includegraphics[width=1.0\linewidth]{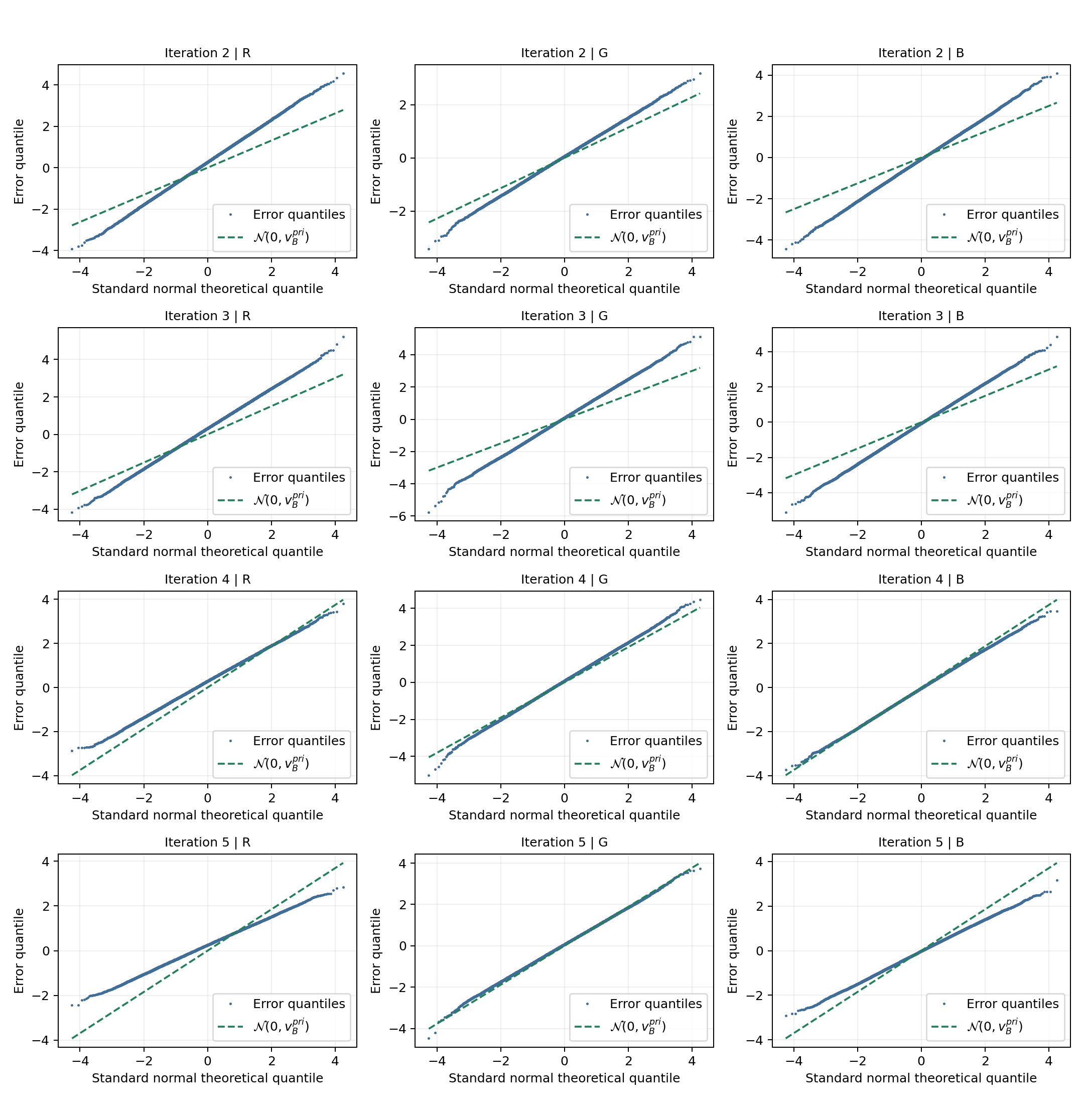}
    \caption{Q-Q plots of the error $\bm x_B^{pri}-\bm x$ at iterations 2, 3, 4 and 5.}
    \label{fig:qq_805}
\end{figure*}

Fig. \ref{fig:qq_805} presents the Q–Q plots of the error $\bm x_B^{pri}-\bm x$ in all three channels at iterations 2, 3, 4 and 5. It can be observed that the empirical error quantiles exhibit an approximately linear relationship with the theoretical Gaussian quantiles. However, some deviations are noticeable in the tails, and a certain discrepancy exists between the empirical standard deviations of the error and the predicted standard deviations $\sqrt{v_B^{pri}}$.
  
  To further assess whether the error distributions satisfy the Gaussian assumption, we have conducted the D’Agostino–Pearson normality test on the error $\bm x_B^{pri}-\bm x$ at iterations 2, 3, 4 and 5. The corresponding results are summarized in Table \ref{tab:normality_test_822}. This test evaluates deviations from normality based on the skewness and excess kurtosis of the empirical distributions, providing additional quantitative evidence to complement the Q–Q plots. A significance level of 0.05 is adopted, with $p$-values below this threshold indicating statistically significant deviations from normality. As shown in Table \ref{tab:normality_test_822}, the normality test results indicate that the error distributions exhibit different degrees of deviation from Gaussianity across channels and iterations. In the early iterations, several channels yield $p$-values below the significance level of 0.05, suggesting statistically significant deviations from normality. 
  % However, as the iterations proceed, the $p$-values generally increase. In particular, from the 21st iteration onward, all three channels yield $p$-values greater than 0.05, indicating that the null hypothesis of normality cannot be rejected. These results suggest that the empirical distributions of the error gradually become more consistent with Gaussian distributions as the iterations proceed.

  Furthermore, to provide additional evidence that the error $\bm x_B^{pri}-\bm x$ does not follow an i.i.d. Gaussian distribution, we train two denoisers with identical architectures but different training datasets. Denoiser-1 is trained on images corrupted by AWGN, whereas Denoiser-2 is trained on the data fed from Module A to Module B during the first iteration of STMP, i.e. ${\bm x_B^{pri}}^{(1)}$. The two denoisers share an identical network architecture and training configuration, with the only difference being their training datasets. We use \(\bm{x}_B^{{pri}}\) as the input to both denoisers, and their reconstruction results are reported in Table \ref{tab:denoiser_comparison}. As can be observed, Denoiser-2 achieves substantially better performance than Denoiser-1, whereas Denoiser-1 almost completely fails to recover the clean images. This pronounced performance gap provides strong evidence of a substantial distribution mismatch between the input data and the training data of Denoiser-1. This observation strongly supports our hypothesis that the elements of the error \(\bm{x}_B^{\mathrm{pri}}-\bm{x}\) do not follow an i.i.d. Gaussian distribution.
\begin{table}[t]
    \centering
    \caption{Performance comparison between two denoisers on FFHQ dataset with a compression ratio of 0.1 and a noise level of 0.05.}
    \label{tab:denoiser_comparison}
    \begin{tabular}{c|ccc}
        \toprule
        Method & PSNR (dB) $\uparrow$ & SSIM $\uparrow$ & LPIPS $\downarrow$ \\
        \midrule
        Denoiser-1 & $9.97$ & $0.0522$ & $0.8496$ \\
        Denoiser-2 & $27.28$ & $0.7635$ & $0.3306$ \\
        \bottomrule
    \end{tabular}
\end{table}

\subsection{Capacity-Matched Comparison}
The refolding operation in LSTMP introduces only 0.127M additional parameters (an increase of 0.19\%) for each layer, yet achieves a substantial 3.13 dB reduction in NMSE at the first layer under hybrid initialization, thereby effectively accelerating convergence in subsequent layers.

\begin{table}[H]
\centering
\caption{capacity-matched comparison. LSTMP-3 corresponds to the same configuration used in the ablation study, while LSTMP* denotes the configuration with an increased number of parameters.}
\label{tab:lstmp_capacity}
\begin{tabular}{lc}
\toprule
Method & NMSE (dB) for layer 1\,/\,3\,/\,5 \\
\midrule
LSTMP-3   & $-21.41$\,/\,$-24.54$\,/\,$-25.10$ \\
LSTMP*  & $-21.49$\,/\,$-24.55$\,/\,$-25.12$ \\
\bottomrule
\end{tabular}
\end{table}
To demonstrate that the performance gains are not merely attributable to the increased number of parameters, we conduct a capacity-matched comparison in Table \ref{tab:lstmp_capacity}. Specifically, we modify \eqref{downsample} as
\begin{equation}
\begin{split}
    & {\bm x_B^{post}}^{(t,i,j+1)} = {\bm x_B^{post}}^{(t,i,j)} \\
    & \quad +\left(\text{Conv}([{\bm x_{B,\uparrow k}^{post}}^{(t,i,j)}, {\bm x_{B,\uparrow k}^{post}}^{(t,i,j)}, {\bm x_{B,\uparrow k}^{post}}^{(t,i,j)}])\right)_{\downarrow k},
    \label{downsample2}
\end{split}
\end{equation}
thereby matching the parameter count of the refolding-based configuration while preventing likelihood information from being incorporated into the denoiser. As shown in Table \ref{tab:lstmp_capacity}, LSTMP-3 corresponds to the same configuration used in the ablation study(i.e., employing deep unfolding and hybrid initialization without refolding). LSTMP* represents the configuration obtained by adding extra parameters through \eqref{downsample2} to the original LSTMP-3 configuration, without introducing additional likelihood information. LSTMP-3 and LSTMP* achieve comparable experimental results, demonstrating the benefit of incorporating likelihood information beyond the effect of the additional parameters.

\end{document}